\documentclass[review]{elsarticle}
\usepackage{subfigure}

\usepackage[varg]{txfonts}
\usepackage{graphicx}
\usepackage{url}
\usepackage{xcolor}

\begin{document}

\begin{frontmatter}
\title{Pulsar Wind Nebulae as a source of the observed electron and positron excess at high energy: the case of Vela-X\\
\vskip .4cm\small{Accepted for publication in Journal of High Energy Astrophysics}}
\author[InfnMIB]{S. Della Torre}
\author[InfnMIB,MIB]{M. Gervasi}
\author[InfnMIB]{P.G. Rancoita\corref{cor1}}
\ead{PierGiorgio.Rancoita@mib.infn.it}
\author[InfnMIB,Insubria,MIB]{D. Rozza}
\author[InfnMIB,Insubria]{A. Treves}

\cortext[cor1]{Corresponding author}

\address[InfnMIB]{INFN Sezione di Milano Bicocca, I-20126 Milano, Italy}
\address[Insubria]{Universit\`a degli Studi dell'Insubria, I-22100 Como, Italy}
\address[MIB]{Universit\`a degli Studi di Milano Bicocca, I-20126 Milano, Italy}

\begin{abstract}
We investigate, in terms of production from pulsars and their nebulae, the cosmic ray positron and electron fluxes above $\sim10$ GeV, observed by the AMS-02 experiment up to 1 TeV. We concentrate on the Vela-X case. Starting from the gamma-ray photon spectrum of the source, generated via synchrotron and inverse Compton processes, we estimated the electron and positron injection spectra. Several features are fixed from observations of Vela-X and unknown parameters are borrowed from the Crab nebula. The particle spectra produced in the pulsar wind nebula are then propagated up to the Solar System, using a diffusion model. Differently from previous works, the omnidirectional intensity excess for electrons and positrons is obtained as a difference between the AMS-02 data and the corresponding local interstellar spectrum. An equal amount of electron and positron excess is observed and we interpreted this excess (above $\sim$100 GeV in the AMS-02 data) as a supply coming from Vela-X. The particle contribution is consistent with models predicting the gamma-ray emission at the source. The input of a few more young pulsars is also allowed, while below $\sim$100 GeV more aged pulsars could be the main contributors.
\end{abstract}


\end{frontmatter}


\section{Introduction}\label{Intro}
The AMS-02 experiment extended the observed cosmic ray (CR) electron, positron and electron plus positron spectra from 0.5 GeV up to 700 GeV, 500 GeV and 1 TeV respectively \citep[see][]{PhysRevLett2,PhysRevLett3}. CR particles generated and accelerated at known sources are considered as primaries. For instance, the main component of electron spectrum is that produced by supernova remnants (SNR). CRs are also produced directly inside the interstellar medium (ISM). In fact, positrons were supposed to be mainly originated from the decay of muons produced by CR interactions with the ISM \citep[e.g.,][]{MS98}. These particles are commonly referred to as secondaries. Primary plus secondary CR spectra outside the region interested by the solar activity (i.e., the heliosphere) are known as local interstellar spectra (LIS, see e.g., Sect. \ref{SecLIS}). In this work we focus our attention on electron and positron spectra. Moreover, we will refer to electrons produced in SNR and in the ISM as the ``classical'' electron LIS and to positrons produced in the ISM as the ``classical'' positron LIS (e.g., ``classical'' LIS, hereafter cLIS).\\
\indent At low energy, less than $\sim10$ GeV, due to solar modulation, the observed CRs spectra deviate from LIS's (see Fig. \ref{FigLIS} and, for instance, \citealt{Bobik2012ApJ,DellaTorrePosFra}). At higher energy, it is commonly acknowledged that, inside the heliosphere, particle propagation is little affected by solar modulation, thus the omnidirectional distribution is the one determined by the LIS. Nevertheless observed spectra of electrons and positrons \citep[see][]{PhysRevLett2,PhysRevLett3} exceed the cLIS computed with GALPROP (see Sect. \ref{SecLIS}) at high energies (e.g., see Fig. \ref{FigLIS}). We evaluated the excess for electrons and positrons subtracting the cLIS from the AMS-02 fluxes and we found an equal amount for the electron excess and the positron one (see Fig. \ref{fig:signal}). Therefore, we take the electron plus positron flux observed by AMS-02 as reference data, extending the comparison with models up to 1 TeV due to the experimental accuracy.\\
\indent In this paper, we investigate possible astrophysical sources of positrons and electrons i.e., pulsars and their nebulae (see Sect. \ref{SecSource} and e.g., \citealt{Linden2013,Yin2013,Yuan2013}), which may account for the flux excess, without the need to look for more exotic explanations, e.g., in the framework of dark matter scenarios \citep[see e.g.,][]{Yuan2013,Ibe2013,Feng2013}. Previous works already explored this scenario (see e.g., \citealt{DiMauro2014,Gaggero2014,Lin2015,Boudaud2015,Serpico2012,Li2014}). Differently from the usual approaches, we evaluated the contribution from astrophysical sources reproducing photon spectra in agreement with the observations.\\
\indent In particular we consider Vela-X as a source of pair production and acceleration. The positron and electron injection spectra are obtained using the diffusion process inside the pulsar wind nebula (PWN) of Vela-X (see Sect. \ref{SecPWN} and e.g., \citealt{Zhang2008}). Shape and luminosity of the flux are fitted in order to be consistent with the gamma-rays spectrum observed from the source. Using a diffusion model described in \citet{Malyshev09}, we evaluated the particle spectra at the Earth position (see Sect. \ref{Sec_Prop}). A comparison with experimental data is, finally, discussed in Sect. \ref{SecRes}, while we report our considerations and conclusions in Sect. \ref{SecDis}.\\
\indent Preliminary results, on materials obtained in this paper, were presented as conference contributions in \citet{DellaTorre:2013opa,RozzaICATPP13}.

\section{Electron and positron Spectra}\label{SecLIS}
\subsection{The ``classical'' Local Interstellar Spectra}
\begin{figure}
\subfigure
{\includegraphics[width=1.\textwidth]{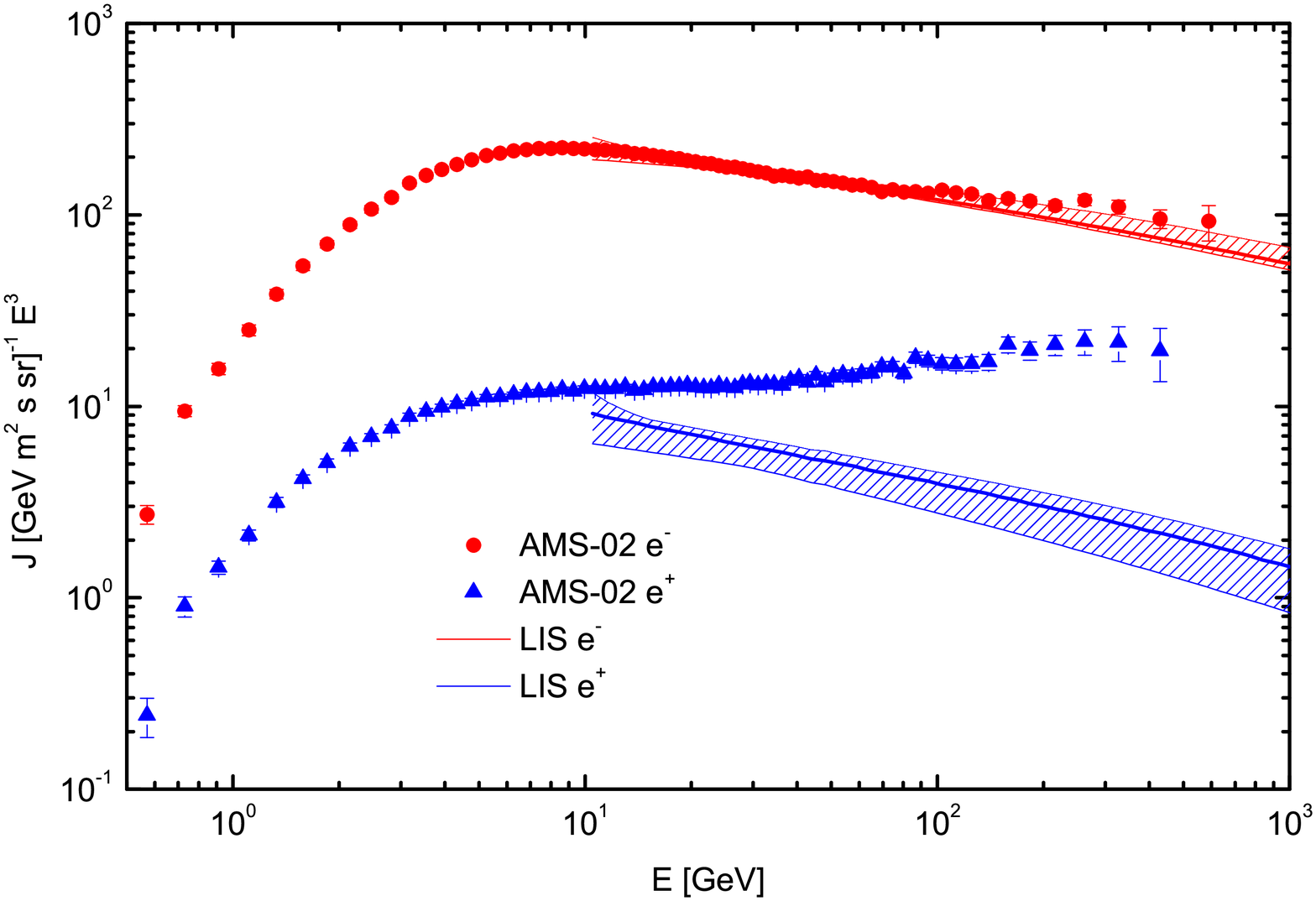}}
\qquad\subfigure
{\includegraphics[width=1.\textwidth]{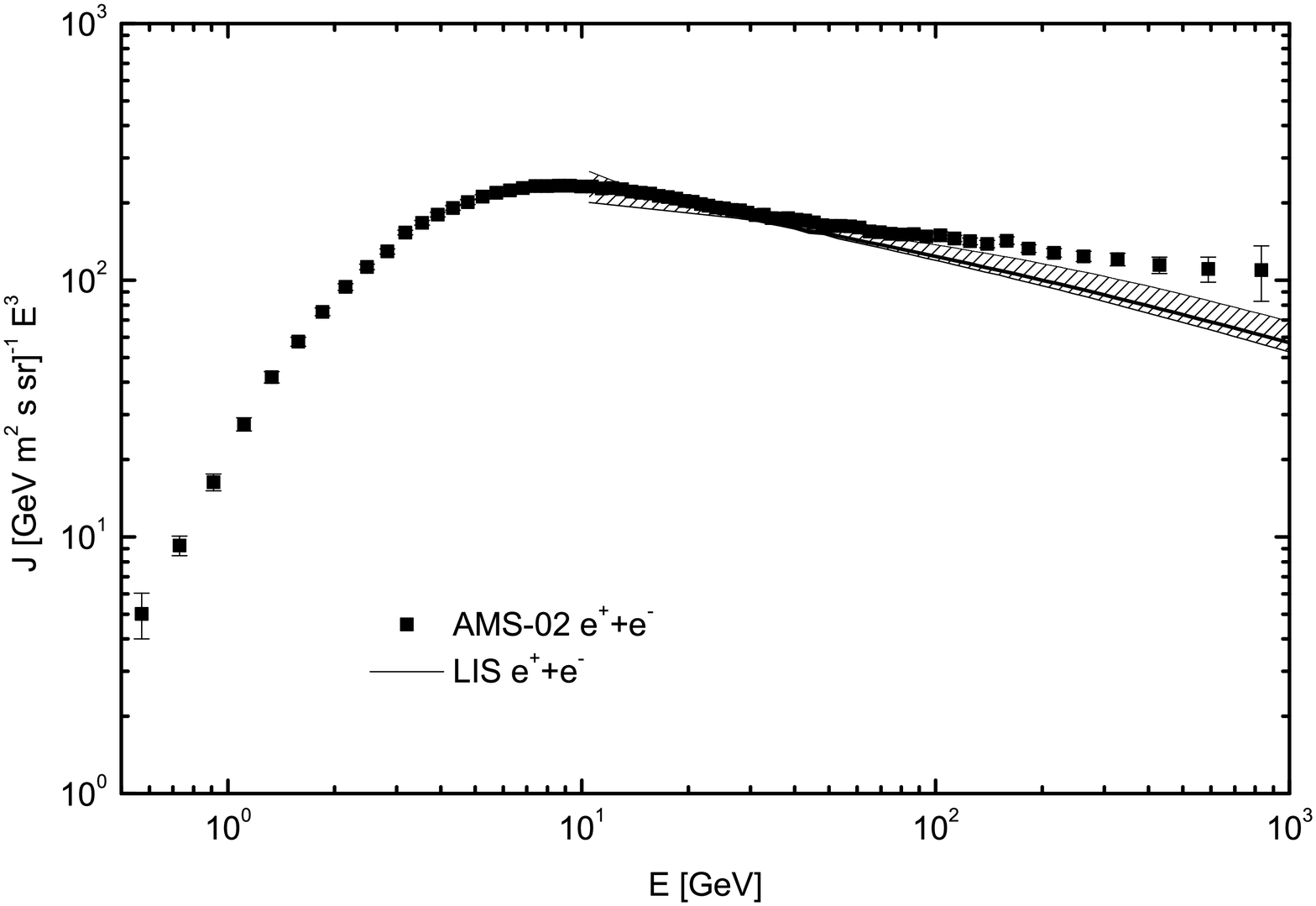}}
\caption{Electron, positron (top) and electron plus positron (bottom) omnidirectional intiensities observed by AMS-02 \citep{PhysRevLett2,PhysRevLett3} and the cLIS's evaluated using GALPROP (obtained with the parameters reported in Tab. \ref{GalPar}) in solid lines; the allowed range of the parameters (reported in Tab. \ref{Tab_GalParRange}) is kept into account by the shadowed bands.}\label{FigLIS}
\end{figure}

The propagation of cosmic rays in the Galaxy can be described by the diffusion equation (e.g., \citealt[Chap.~3]{Ginzburg64} and \citealt{Ginzburg76}):
\begin{equation}\label{EqDiffCR}
 \frac{\partial n_{i}}{\partial t} = Q_{i} + \vec{\nabla}\cdot\left[D_{i}\,\vec{\nabla}n_{i}\right] + \frac{\partial}{\partial E}\left[b_{i}\,n_{i}\right] - p_{i}n_{i} + P_{i},
\end{equation}
where the time evolution of the energy density $n_{i}=dN_{i}/dE$ of cosmic ray species $i$ with energy $E$ depends on the source term $Q_{i}$, diffusion coefficient $D_{i}$ usually described by a power law in the energy $D(E)=D_{0}(E/E_{0})^{\delta}$, \citep[see][]{Ptuskin06,Galprop07}, the change of the particle energy per unit time $b_{i}$, catastrophic processes $p_{i}$ and nuclei collisions $P_{i}$. Equation (\ref{EqDiffCR}) accounts for i) the propagation of primary components like, e.g., electrons, protons and carbon nuclei mainly accelerated in SNRs \citep[Chap.~4]{Ginzburg64} and ii) the production of secondary spectra like, e.g., positrons and Boron nuclei produced from interaction of primary CRs with the ISM.\\
\indent The most recent data provided by PAMELA and AMS-02 were discussed by many authors. For instance, \citet{Blasi2012} introduced two slopes in the diffusion coefficient in different energy range to explain the proton and helium spectrum; \citet{Shaviv2009} interpreted the CR spectra using an inhomogeneity source distribution following the stars concentration in the galactic spiral arms; \citet{Blum2013} modelled the rising of the PAMELA and AMS-02 positron flux as due to purely secondary origin, without taking into account the energy losses. The GALPROP model solves numerically Eq.\ (\ref{EqDiffCR}) for all the relevant CR species in a cylindrically symmetric space \citep{vladimirov}, with a galactic radius $R_{Gal.}$ and height $h_{Gal.}$. Therefore, hereafter, we will use the most comprehensive propagation model; i.e., the GALPROP model by which we evaluate, at the same time and with the same propagation parameters, the local interstellar spectra of several kind of particles: protons, electrons, ions, anti-particles and photons used for the electron energy loss.\\
\indent The GALPROP code\footnote{http://galprop.stanford.edu/webrun.php} returns the predicted ``classical'' LIS for the specific particle at the Solar System. The solution of Eq.\ (\ref{EqDiffCR}) depends on parameters like the boundary conditions of the galactic effective volume for CRs diffusion, the diffusion coefficient and the injection spectra characterized by power laws with different spectral indices for nuclei, protons ($\gamma_p$) and primary electrons ($\gamma_e$). To determine these parameters, we compared the spectra obtained in this way with the experimental data above $\sim10$ GeV (energies high enough to neglect solar modulation effects), then we tuned the coefficients minimizing the discrepancies. The calculated spectra were normalized at 50 GeV with measured proton, electron and ion fluxes at Earth. For proton and electron spectra we used the AMS-02 data \citep{PhysRevLett2,SadaICRC}, while for the ions ratios B/C, Be/B, Be/C, Li/B, Li/Be and Li/C we referred to the online cosmic ray database reported in \citet{Maurin2013CRDB}.\\
\indent The available data are best described using the parameters listed in Tab. \ref{GalPar}. In Fig.\ \ref{FigLIS} we reported the comparison between the cLIS's and AMS-02 data for electron, positron and electron plus positron spectra. The cLIS's, coming from GALPROP, were reported, in solid lines, for energy above 10 GeV where solar modulation is negligible.\\
\begin{table}
\centering
\begin{tabular}{|c|c|}
 \hline
 Parameter & Value \\
 \hline
 $R_{Gal.}$ & $30$~kpc \\
 $h_{Gal.}$ & $\pm4$~kpc \\
 $D_{0}$ & $5.8\cdot10^{28}$~cm$^{2}$~s$^{-1}$ \\
 $\delta$ & $0.33$ \\
 $E_{0}$ & $4$~GeV \\
 $v_{A}$ & $30$~km~s$^{-1}$ \\
 $\gamma_{p}$ & $1.98$ ($E<9$~GeV), $2.42$ ($E>9$~GeV) \\
 $\gamma_{e}$ & $1.7$ ($E<4$~GeV), $2.68$ ($E>4$~GeV) \\
 \hline
\end{tabular}
\caption{Propagation parameters used in GALPROP code to determine the ``classical'' electron and positron LIS's.}\label{GalPar}
\end{table}
\indent Since the choice of the values of Tab. \ref{GalPar} is not unique, we modified, one by one, the main GALPROP parameters responsible for the diffused spectra (the galactic height and the diffusion coefficient), as reported in Tab. \ref{Tab_GalParRange}. The ranges of this parameters were determined keeping the produced LIS within the experimental data errors, with special regard to the Boron over Carbon ratio \citep[see e.g., ][]{BCAMS1,BCCream,BCPAMELA,BCTracer,TingAMSDays2015}. This overall uncertainty is included in the shadowed regions of Fig.\ \ref{FigLIS}. We also found that the cLIS, obtained with the parameters of Tab. \ref{GalPar} and \ref{Tab_GalParRange}, is compatible with the $\bar{p}/p$ ratio of PAMELA \citep{Adriani2013} and AMS-02 \citep{TingAMSDays2015}.
\begin{table}[htbp]
\centering
\begin{tabular}{|c|c|}
 \hline
 Parameters & Range \\
 \hline
 Galactic height (kpc) & $2<h_{Gal.}<6$ \\
 Diffusion Coefficient Constant (cm$^{2}$s$^{-1}$) & $4\cdot10^{28}<D_{0}<10^{29}$ \\
 Diffusion Coefficient Index & $0.3<\delta<0.4$ \\
 \hline
\end{tabular}
\caption{Ranges of propagation parameters used in GALPROP code to determine the errors in the LIS evaluation.}\label{Tab_GalParRange}
\end{table}

\subsection{Electron and positron flux excess at high energy}\label{SecSignal}
\begin{figure}\resizebox{\hsize}{!}{\includegraphics[width=0.8\textwidth]{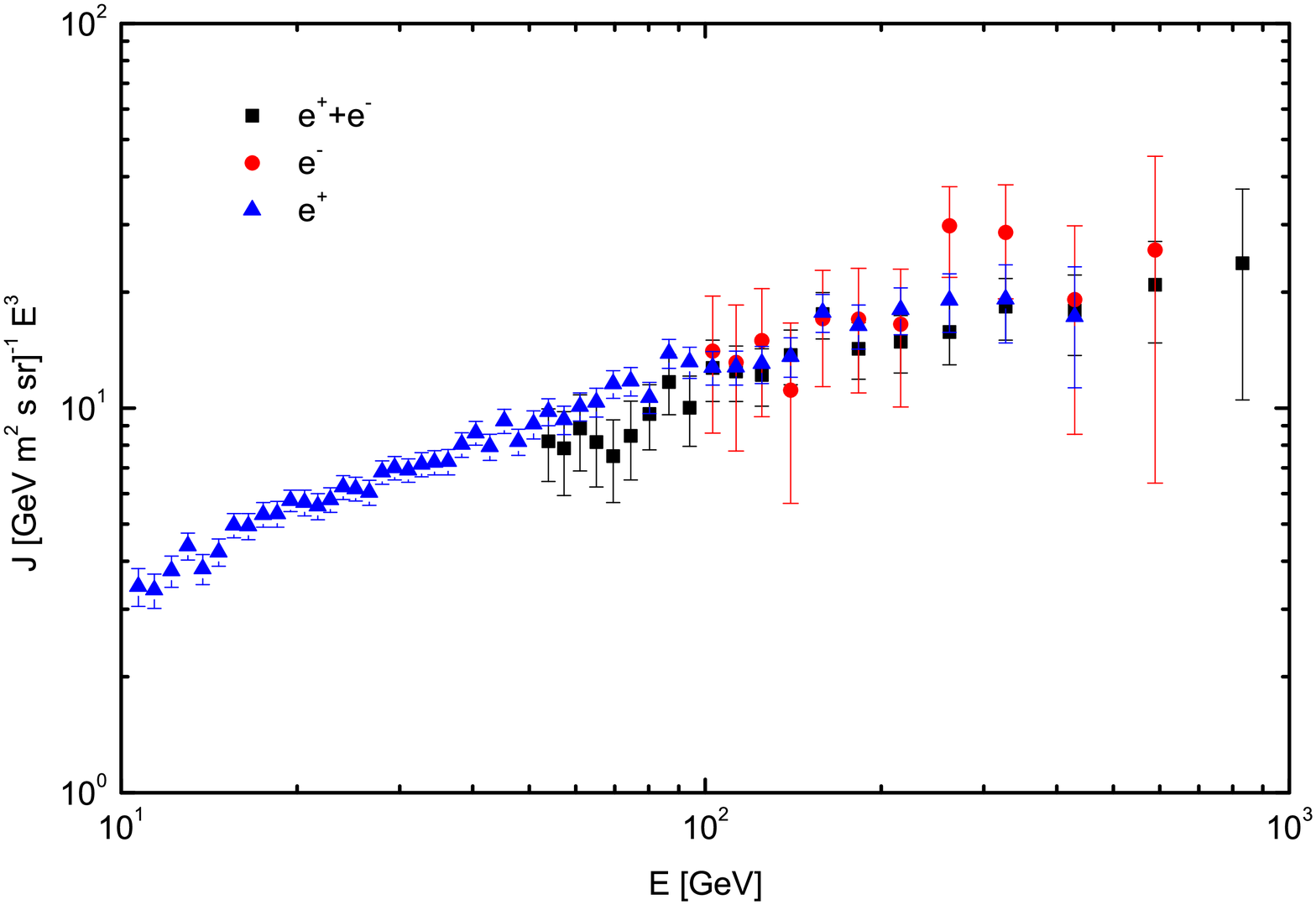}}
 \caption{Omnidirectional intensity excess for electrons, positrons and half of all electron, obtained as a difference between the AMS-02 flux and the corresponding ``classical'' LIS.}
 \label{fig:signal}
\end{figure}

The omnidirectional intensity excess for electrons and positrons are shown in Fig.\ref{fig:signal}.

The difference between the observed AMS-02 spectra and GALPROP cLIS's (solid lines of Fig. \ref{FigLIS}) were calculated for energy above $\sim10$ GeV (where the solar modulation effects are negligible) and requiring at least a difference (above 10\%) between the two fluxes. Under these constraints the electron and positron signal is reported for energy above 90 GeV and above 10 GeV, respectively. We report also the electron plus positron spectrum, above 50 GeV, divided by a factor two for a comparison with respect to the other data. The error bars of these data come from the experimental observations. We can remark how these excess spectra of positrons and electrons can be fitted using similar power laws. The electron signal spectral index, resulting from the fit, is $-(2.503\pm0.353)$, for positrons we have $-(2.502\pm0.030)$, while for electron plus positron spectrum we have $-(2.568\pm0.088)$. The points of Fig. \ref{fig:signal} are dependent on the parameters used in GALPROP. The uncertainties due to the choice of the GALPROP parameters result as a scale factor of the omnidirectional intensities in Fig. \ref{fig:signal}. This uncertainty is mostly constrained by the positron spectrum and can be accounted as a scale factor of $\sim5\%$ at 100 GeV and above, while at lower energy it is $\leq20\%$. For the electron plus positron spectrum, the uncertainty at 1 TeV is about 25\%. Hereafter, we will compare our source models with the electron plus positron spectrum since the upper point reaches 1 TeV.

\section{Pulsars as possible sources of the excess components}\label{SecSource}
Looking at Fig. \ref{fig:signal}, above 50 GeV, a region marginally affected by the solar modulation, it is possible to say that the electron and positron signals are compatible. Thus, the dominant physical process is expected to proceed via a pair production mechanism. Pulsars (PSRs) are among the most likely sources of electron-positron pairs. The high magnetic field and the fast rotation of these neutron stars lead to huge electric fields \citep{Goldreich69}. Electrons produce high energy photons via curvature radiation. The interactions between photons and high magnetic fields generate pairs \citep{Sturrock71} that can generate again curvature photons. At the end, electromagnetic showers are produced.\\
\indent PWN identify the region around the pulsar where a relativistic magnetized wind is populated with electrons and positrons \citep[e.g.,][]{Harding2004,Blasi2010}. PWN are widely believed to be responsible for the acceleration of cosmic rays up to energies of $10^{15}$ eV \citep[e.g.,][]{Rees1974,Kennel1984694,Kennel1984710}. The central pulsar converts its spin-down power into a relativistic wind injected near the magnetosphere. The electrons and positrons in the wind, interacting with shock fronts, are accelerated and get a power-law energy spectrum. They radiate at lower energies, from radio frequencies to X-rays, through the synchrotron process in the magnetic field of the nebula. The highest energy part of the spectrum comes from inverse Compton scattering of the radiation field: synchrotron radiation generated by the PWN, cosmic microwave background (CMB), infrared and star-light photons (e.g., \citealt{deJager1992,Atoyan1996,Hillas1998}). The gamma ray spectra of PWN reach very high energy, for instance, tens of TeV. The observed high-energy emission from the Crab Nebula has been modelled in detail by several authors \citep[see][]{deJager1992,Atoyan1996,deJager1996,Hillas1998,Bednarek2003}. High-energy processes in other PWN such as that of the Vela-X nebula, the nebulae around PSR 1706-44, PSR 1509-58, 3C 58, CTB 80 and other few nebulae have been also studied in \citet{duPlessis1995,Aharonian1997,Sefako2003,Bednarek2003,Horns2006}. It was also suggested that the production of gamma-rays in the interactions of hadrons with the matter of the supernova could contribute to the higher energy end of the observed spectrum, especially in the case of the youngest nebulae \citep[see][]{Cheng1990,Atoyan1996,Bednarek1997,Horns2006}.\\

\subsection{Electron and positron spectra at the source}\label{SecPWN}
Since pulsars are rotating magnetized neutron stars, basically PSR lose their energy through the electromagnetic radiation of the magnetic dipole spinning around a tilted axis. Within a simplified model, it is possible to determine the main physical parameters of the pulsar (like magnetic field or energy loss) starting from the rotation frequency ($\nu$) of the pulsar, its first ($\dot{\nu}$) and second ($\ddot{\nu}$) derivative. The spin down of the pulsar is assumed as:
\begin{equation}\label{Eq_nudot}
\dot{\nu}=-k\nu^n,
\end{equation}
where $k$ depends on the magnetic moment and on the moment of inertia of the neutron star and $n$ is the braking index that would be equal to 3 for a magnetic dipole. Assuming constant $k$, from Eq. (\ref{Eq_nudot}) we get:
\begin{equation}\label{Eq_n}
n=\frac{\nu\ddot{\nu}}{\dot{\nu}^2}.
\end{equation}
Moreover, integrating Eq. (\ref{Eq_nudot}), assuming constant $k$, the age of the pulsar ($\tau_{age}$) can be obtained:
\begin{equation}\label{Eq_tauage}
\tau_{age}=-\frac{\nu}{(n-1)\dot{\nu}}\left[1-\left(\frac{\nu}{\nu_0}\right)^{n-1}\right].
\end{equation}
Here $\nu_0$ is the birth rotation frequency. The time evolution of spin-down luminosity is given by \citep[e.g.,][]{Martin2012}
\begin{equation}\label{Eq_L}
L(t)=L_0 \left(1 + \frac{t}{\tau_0}\right)^{-\frac{n+1}{n-1}}
\end{equation}
where $L_0$ is the initial spin-down luminosity and $\tau_0 = P_0/(n-1)\dot{P_0}$ is the initial spin-down time scale.\\
To evaluate the energy spectrum of electrons and positrons at the source, we follow the approach described in \citet{Zhang2008}. This method is similar with others reported in \citet{Tanaka2010,Martin2012}. The time dependent model by \citet{Zhang2008} describes both particle and photon injection spectra from the nebula. It assumes an initial particle injection rate, produced by the pulsar, that follows a broken power law with indices $\alpha_1$ and $\alpha_2$ and energy break $E_b$:
\begin{equation}\label{Eq_QE}
 Q(E_e,t)=\left\{\begin{array}{ll}
	    Q_0(t)(E_e/E_b)^{\alpha_1} & \textrm{if $E_e<E_b$}\\
	    Q_0(t)(E_e/E_b)^{\alpha_2} & \textrm{if $E_e>E_b$}\\
      \end{array}\right.,
\end{equation}
where $E_e$ is the particle kinetic energy and $Q_0(t)$ can be derived by requiring the continuity of the two power laws and that
\begin{equation}
 \int{Q(E_e,t)E_edE_e}=\eta L(t),
\end{equation}
with $\eta$ the conversion factor of the spin-down power $L(t)$ into particle luminosity. The diffusion equation for the differential electron density $n_e(E_e,t)=dN(E_e,t)/dE_e$ can be approximated as (\citet{Zhang2008,Tanaka2010,Martin2012}):
\begin{equation}\label{Eq_diffEq}
 \frac{dn_e(E_e,t)}{dt}=Q(E_e,t)-\frac{n_e(E_e,t)}{\tau_{syn}(t)}-\frac{n_e(E_e,t)}{\tau_{esc}(t)}.
\end{equation}
The particle spectrum obtained as a solution of the Eq. (\ref{Eq_diffEq}) over time from $t=0$ to $t=T$ (age of the PWN) is:
\begin{equation}\label{Eq_SoldN}
 \frac{dN(E_e,T)}{dE_e}=\int_0^T{Q(E_e,t)\exp\left(-\frac{T-t}{\tau_{eff}}\right)dt},
\end{equation}
where $\tau_{eff}^{-1}=\tau_{syn}^{-1}+\tau_{esc}^{-1}$ corresponding to the lifetime of an electron with respect to both synchrotron energy loss and escape timescale (i.e., the time to diffuse 1 PWN radius). In Ref. \citet{Zhang2008}, the authors applied their model to the Crab nebula, for which an accurate spectral energy distribution is known, due to numerous observations from radio frequency up to gamma rays (see \citealt{Zhang2008} and references therein). The low energy spectral index ($\alpha_1$ in eq.\ref{Eq_QE}) is generally fixed by fitting the synchrotron spectrum at low frequency. The high energy spectral index ($\alpha_2$ in eq.\ref{Eq_QE}) is however related to the inverse Compton process. In the next subsection we will use this result as source spectrum of the electron and positron excess spectra at Earth.

\subsection{Electron and positron spectra at the Earth}\label{Sec_Prop}
Following the approach reported in \citet{Malyshev09}, the time evolution of the energy density $n_{e}(\vec{x},E,t)$ of electrons or positrons from a single source distant $\vec{x}$ from the Solar System, with energy $E$ and after a diffusion time $t$, is obtained disregarding the last two terms of Eq.\ (\ref{EqDiffCR}) \citep[see][]{Delahaye2010}, i.e.,
\begin{eqnarray}\label{EqDiff}
 \frac{\partial n_{e}(\vec{x},E,t)}{\partial t} &=& Q_e(E,t) + \vec{\nabla}\cdot\left[D(E)\vec{\nabla}n_{e}(\vec{x},E,t)\right]\nonumber\\
 & & + \frac{\partial}{\partial E}\left[b(E)n_{e}(\vec{x},E,t)\right].
\end{eqnarray}
In Eq.\ (\ref{EqDiff}), the term $b(E)$ accounts for the rate of energy loss due to ionization, Bremsstrahlung, synchrotron and inverse Compton processes \citep[e.g.,][Chap.~4]{Schlickeiser}. However, above $\sim1$ GeV, the only relevant mechanisms are synchrotron and inverse Compton. Furthermore, above few GeV, using an average interstellar magnetic field of $3\ \mu$G and the photon radiation fields reported in \citet{Delahaye2010} (Table 2, model M1), the fit of the total energy loss rate can be described by a power law as in:
\begin{equation}\label{EqbE}
 \frac{dE}{dt}=-b(E)\sim-b_0E^2,
\end{equation}
where $b_0\sim7\cdot10^{-17}$ GeV$^{-1}$s$^{-1}$ (value in agreement with those reported in e.g., \citealt{Kobayashi2004-0308470,Atoyan1995}). Due to the high rate of energy loss, a positron or an electron of 100 GeV dissipates most of its energy in about $10^6$ years and can diffusively travel up to a typical distance of about 2 kpc. Thus, sources responsible of the high energy positron and electron excess (observed by PAMELA, \citealt{Pamelaposfrac}, and AMS-02, \citealt{AMS02posfrac,PhysRevLett1}) are located in a region relatively close to Earth (within a distance of $\sim2$ kpc).\\
\indent Then, we fit the injection spectrum, i.e., the results of Eq. (\ref{Eq_SoldN}), as a function of the initial energy $E_0$, using a power law with  spectral index $\alpha$ and an exponential energy cut-off $E_{cut}$:
\begin{equation}\label{EqQE}
 Q_e(E_0,t)=Q_{e,0}(t)E_0^{-\alpha}\exp\left(-\frac{E_0}{E_{cut}}\right),
\end{equation}
the interstellar diffused spectra of electrons and positrons from Eq. (\ref{EqDiff}) is \citep{Ginzburg76,Malyshev09}:
\begin{eqnarray}\label{flux}
J(\vec{x},E,t) &=& \frac{c}{4\pi}n_{e}(\vec{x},E,t)\nonumber\\
&=&\frac{c}{4\pi}\frac{Q_{e,0}}{(4\pi\lambda_d^2)^{3/2}}E^{-\alpha}\left(1-b_0tE\right)^{\alpha-2}\nonumber\\
& & \times\exp\left[-\frac{E}{E_{cut}(1-b_0tE)}\right]\exp\left(-\frac{|\vec{x}|^{2}}{4\lambda_d^2}\right),
\end{eqnarray}
where $t$ is the diffusion time (i.e., the time spent to reach the solar system), $\vec{x}$ is the distance between the source position and the Earth and $\lambda_{d}$ is the mean distance travelled by particles with initial energy $E_0=E/(1-b_0tE)$ down to energy $E$ resulting from both energy loss and diffusion processes given by
\begin{equation}\label{lambda}
\lambda_{d}(E,E_{0})=\left(\int^{E_{0}}_{E}\frac{D(E')dE'}{b(E')}\right)^{1/2}.
\end{equation}
Eq. (\ref{flux}) allows one to evaluate the electron and positron spectra at Earth coming from PWN for energies beyond $\sim10$ GeV, i.e., energies high enough to neglect solar modulation effects.

\section{The case of Vela-X: Results}\label{SecRes}
The pulsar J0534+2200, located inside the Crab Nebula, the remnant of a supernova explosion occurring in A.D. 1054, is an extremely well studied object. This young source can give us information about the first step of the life of a generic PSR/PWN. Note that the Crab distance is about 2 kpc and, due to the age, we can not see yet particles coming from that source at Earth. The TeVCat catalogue\footnote{http://tevcat.uchicago.edu/} contains less than 40 PWN observed in the TeV energy range. Only five of them are closer than 2 kpc and were observed by Cherenkov telescope experiments like HESS \citep{HessCrab,HESS06}, Veritas \citep{VeriCTA2013,VeriBoom2009} and Magic \citep{Magic2012}. Vela-X belongs to this sample. These observations regard a small fraction of known pulsars and they are much less complete and accurate in comparison with the Crab Spectral Energy Distribution (SED). Indeed it is widely believed that PWN are no more observable after the early phase of expansion. \citet{Malyshev09} suggested that all the pulsars have an initial stage as PWN and the lifetime of these objects is about 10$^3$-10$^4$ years. During this phase electrons and positrons are trapped inside the PWN, but later, after a time $T$ from the SN explosion, they are free to propagate. Mature pulsars, like Geminga and Monogem, have no more gamma-ray emission from the nebula, but the electrons and positrons released are still coming to the Earth. For all the older pulsars we do not have information regarding the nebula photon spectrum, the braking index or the birth frequency. For what concerns the PSR age, we can roughly estimate the minimum characteristic age $\tau_{age,c}=-\nu/2\dot{\nu}$ as reported in \citet{Abdo2013}.

\subsection{Model 1: using observed parameters for Vela-X}
Vela-X was detected by HESS \citep{HESS06} in the very high energy gamma ray band, the spectrum can be fitted by a power law with the photon index $\Gamma_{\gamma}=1.45\pm0.09_{stat}\pm0.2_{sys}$ in the energy range between 550 GeV and 65 TeV and an exponential cut-off at an energy of $13.8\pm2.3_{stat}\pm4.1_{sys}$ TeV. The X-ray part was detected using ROSAT combined with ASCA data \citep{VelaX_1995,VelaX_1997}. The spectrum observed in this region has a spectral index of $\Gamma_X\sim2$. The observation of the timing property of the pulsar gives us the basic information to evaluate the braking index as in Eq. (\ref{Eq_n}). Vela is one of the few pulsars \citep{Yue2007} for which it is possible to measure the braking index: $n=1.4\pm0.2$ \citep{Lyne1996}. Therefore, we built a model based on the observed parameters of Vela and we assumed a value of $n$ not dependent on time. The birth period $P_0$ can be evaluated by Eq. (\ref{Eq_tauage}) knowing the frequency, its derivatives and the age of the object. The CRAB source is the only one for which all these parameters are known. For this reason, we assumed the same initial rotation period of the Crab pulsar ($P_0\sim20$ ms, \citealt{Manchester1977}). In this way, Eq. (\ref{Eq_tauage}) gives an age for Vela of about 26 kyrs (instead of the common characteristic age $\tau_{age,c}=11$ kyrs). The results reported in Fig. \ref{Fig_PSR} and \ref{fig:All_LIS10TeVall} (Model 1) come from this analysis. We get a photon spectrum compatible with the HESS and ASCA data requiring a conversion efficiency of the spin down luminosity of about $\eta=0.5\%$ for both electrons and positrons. The Model 1, reported in the figures, is the diffused spectrum at the Earth for which we set $T\sim10$ kyrs, in comparison with the initial spin-down time scale which is evaluated to be $\tau_0\sim29$ kyrs. We also decided to vary the $P_0$-value in the range: $10<P_0\:\textrm{(ms)}<30$. Consequently, Vela age changes from 33 to 19 kyrs. In all cases, we fit the photon spectrum in agreement with the ASCA and HESS data. For this purpose we change the efficiency inside the range $0.25<\eta\:\textrm{(\%)}<1$ respectively and the released time from 22 to 5 kyrs. The band of the model reflects the uncertainty on the Vela pulsar distance that is about 6\% (Vela distance is $287^{+19}_{-17}$ pc) \citep{Abdo2013} and the variation due to the assumed initial rotation period $10<P_0\:\textrm{(ms)}<30$.

\subsection{Model 2: using Crab-like parameters for Vela-X}
\begin{figure}\resizebox{\hsize}{!}{\includegraphics[width=0.8\textwidth]{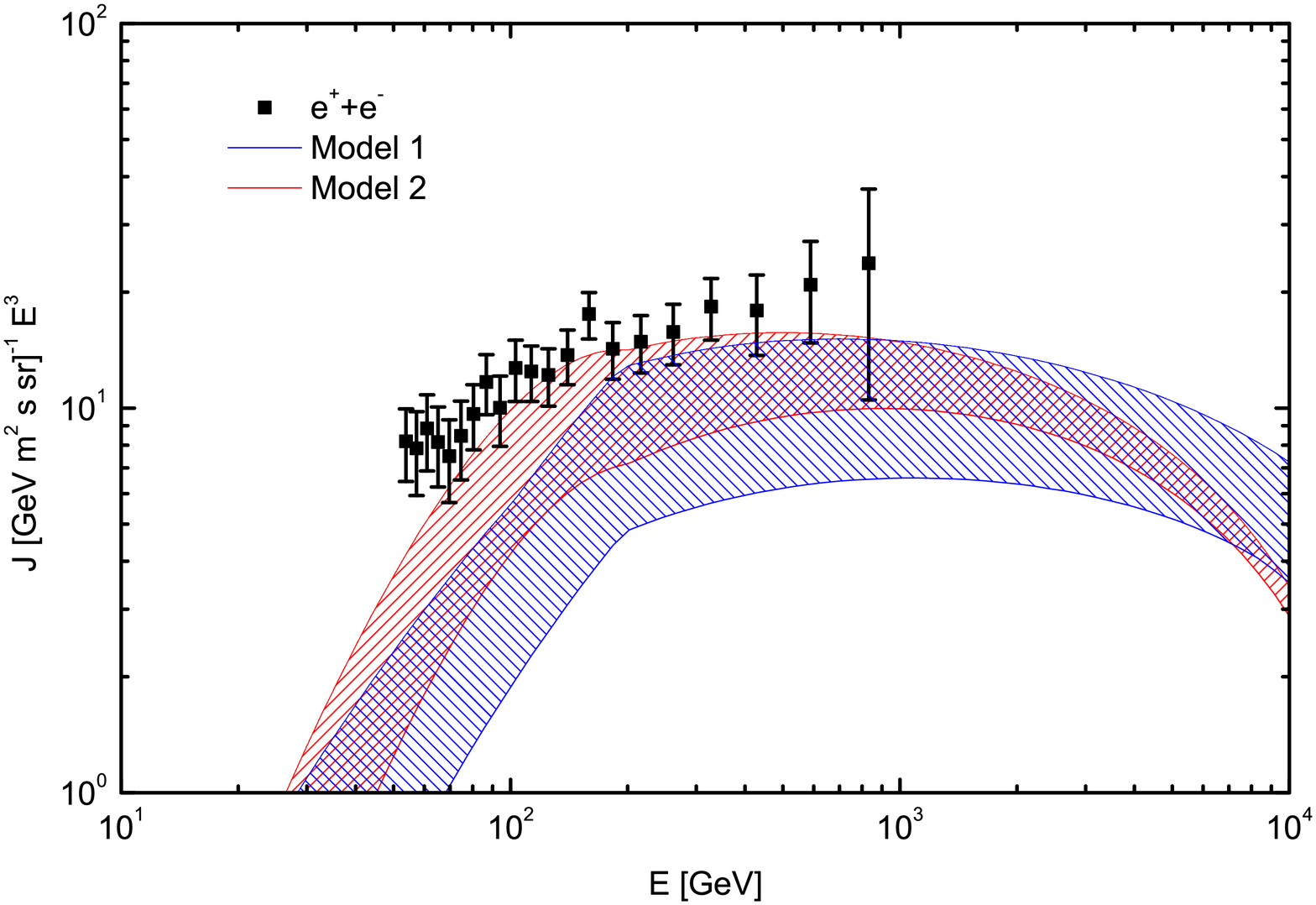}}
 \caption{Two model of this analysis of the expected positron or electron omnidirectional intensities from Vela-X compared with the half electron plus positron signal as in Fig. \ref{FigLIS}. The shadowed bands reflect the uncertainty on Vela distance, that is about 6\%, and the variation due to the assumptions on the initial rotation period.}
 \label{Fig_PSR}
\end{figure}

\begin{figure}
\subfigure
{\includegraphics[width=1.\textwidth]{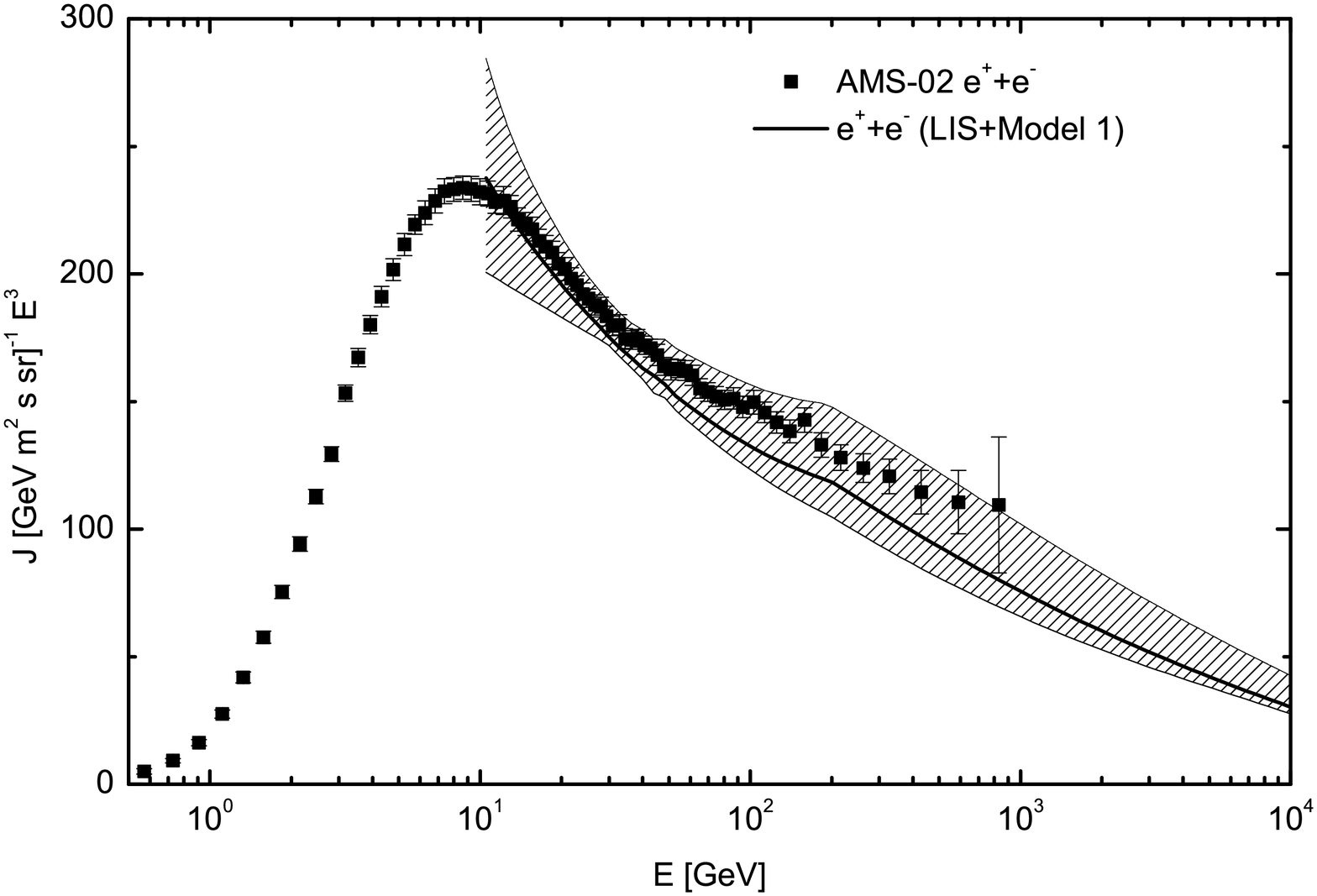}}
\qquad\subfigure
{\includegraphics[width=1.\textwidth]{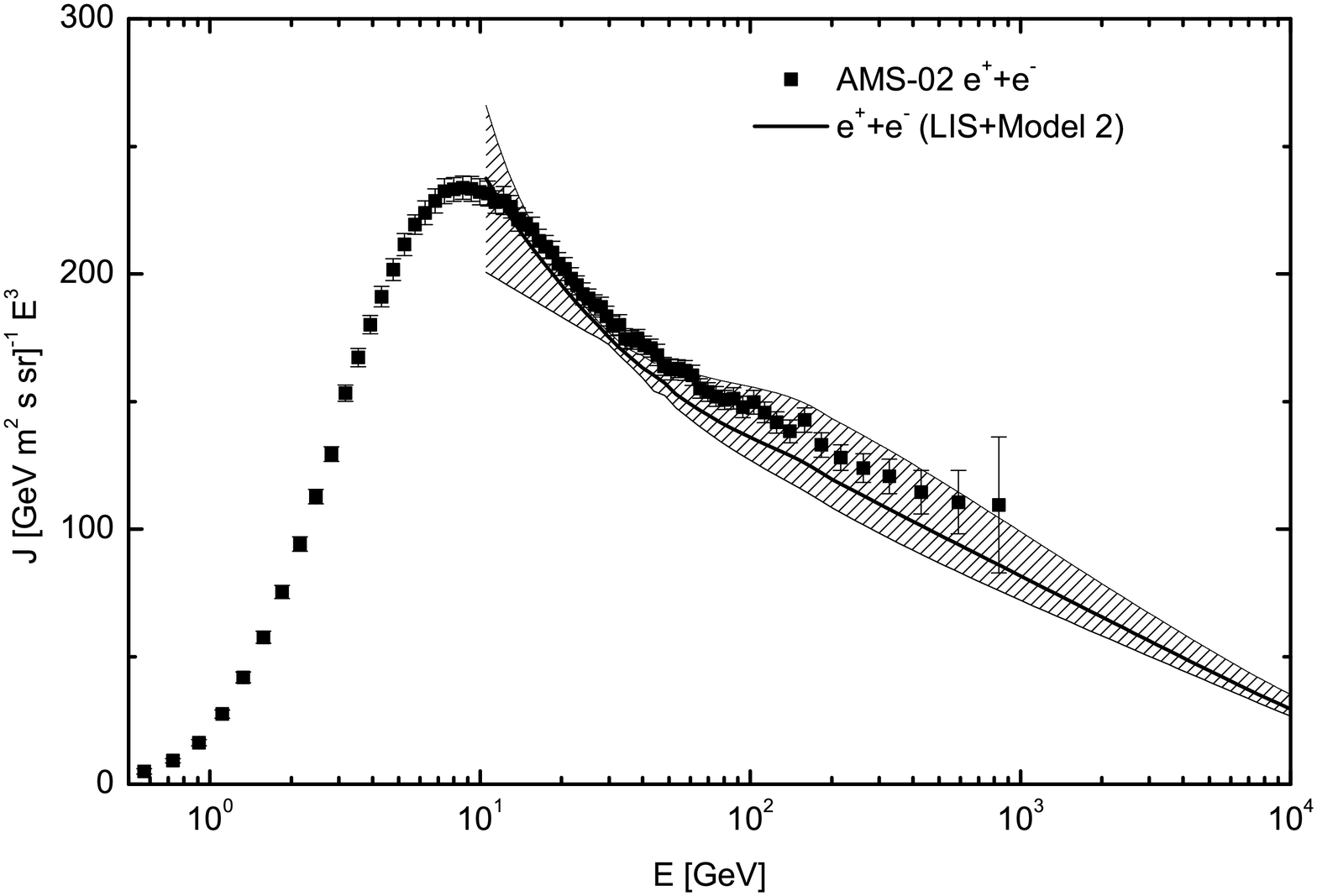}}
\caption{Electron plus positron contribution from cLIS and Vela-X Model 1 and 2 (solid lines) compared with the AMS-02 flux. The bands keep into account the uncertainties of the cLIS parameters (see Tab. \ref{Tab_GalParRange}), the error on the Vela-X distance and, for Model 1, also the variation due to the assumptions on the initial rotation period.}\label{fig:All_LIS10TeVall}
\end{figure}

Observation of the timing properties of the other five pulsars, younger than Vela, gives for the braking index $n$ values in the range between 2 and 3 \citep{Yue2007}. All of them are more similar to the value for the Crab nebula with respect to Vela-X. Therefore, we can alternatively assume that all the pulsars are similar at their birth and then we can take the properties (initial rotation period, breaking index) of Crab. We need to assume that there is a variation of the braking index from 2.5, at the birth, down to 1.4, at later time. It could be associated to some changes in the structure of the neutron star, as suggested by the observation of glitches in the rotation period \citep{Lyne1996}. We do not have a model for this variation and, therefore, it is not currently possible to evaluate the photon spectra at the present day, but we can use the Crab photon spectra observed after $\sim$ 1000 years from the birth. For the Crab-like source the initial spin-down time scale is $\tau_0\sim700$ years and the characteristic age is $\tau_{age,c}\sim11$ kyrs. The conversion efficiency of the spin down power into particle luminosity ($\eta$) is defined by \citet{Zhang2008} ($\eta_{e^+}+\eta_{e^-}=2\eta_{e^+}=0.1$). Therefore, we take electron-positron spectra, normalized to the photon emission like in the Crab nebula, and propagate the source spectrum after $T\sim1000$ years, as assumed in \citet{Malyshev09}. In Fig. \ref{Fig_PSR} and \ref{fig:All_LIS10TeVall} the positron or electron spectrum, obtained from Eq. (\ref{flux}) using all the Crab parameters, is shown for Vela-X (Model 2). The main parameters of the two models are summarized in Tab. \ref{ModelPar}.\\
\begin{table}
\centering
\begin{tabular}{|c|c|c|}
 \hline
 Parameter & Model 1 & Model 2\\
 \hline
 $\alpha_1$ & 1.9 & 1.5\\
 $\alpha_2$ & 2.8 & 2.4\\
 $E_b$ (MeV) & $1.5\cdot10^5$ & $1.5\cdot10^5$\\
 $n$ & 1.4 & 2.5\\
 $\eta$ (\%) & 0.25 - 1 & 5\\
 $P_0$ (ms) & 10 - 30 & 20\\
 $\tau_{age}$ (kyr) & 33 - 19 & 11\\
 $\tau_{0}$ (kyr) & 22 - 35 & 0.7\\
 $T$ (kyr) & 5 - 22 & 1\\
 \hline
\end{tabular}
\caption{Parameters used in Model 1 and 2 for Vela-X nebula.}\label{ModelPar}
\end{table}

\indent From an inspection of Fig. \ref{Fig_PSR}, one may remark that the measured electron and positron intensities can be accounted for by the flux expected from Vela-X within models 1 and 2. Fig. \ref{fig:All_LIS10TeVall} shows the comparison between our models (cLIS and Vela-X contributions) and the absolute AMS-02 electron plus positron flux. The shadowed bands keep into account the uncertainties due to the cLIS parameters (see Tab. \ref{Tab_GalParRange}), the error on Vela-X distance and, for Model 1, also the variation due to the assumptions on the initial rotation period. In the top graph, the band is in agreement with the AMS-02 data, thus, Vela-X alone can account for the electron and positron excess components. In the bottom one, for energies between 20 and 80 GeV, the AMS-02 data are slightly above the model. This discrepancy can be accounted for introducing aged pulsars contributing with lower energy electrons and positrons (in comparison with Vela-X), while the photon emission of their PWN is no more observable. The main contribution coming from these mature pulsars is due to Monogem. Its electron and positron contributions, evaluated with Model 2 (Monogem treated as Crab-like), is a factor 2 higher than the Vela-X one at 40 GeV.
\section{Discussion and Conclusions}\label{SecDis}
In the present paper, we analysed the AMS-02 electron, positron and electron plus positron spectra in the energy range above $\sim10$ GeV. The excess spectra for electrons and positrons were obtained by subtracting the expected ``classical'' LIS's, computed with GALPROP, to the omnidirectional distributions observed by AMS-02. A comparison above $\sim50$~GeV indicates for electrons and positrons the same flux and the same slope. These excess spectra can be accounted for by pulsar sources in which generated electron-positron pairs will be accelerated by the surrounding pulsar wind nebula. In particular, we evaluated electron and positron spectra generated in the Vela-X PWN and propagated them to the Earth. We used two different models built using observed parameters of Vela and Crab nebulae. Finally, we compared results with observations. Both models, taking into account uncertainties and assumptions, are not in disagreement with the AMS-02 excess components at energy larger than $\sim$ 100 GeV. Vela-X Model 2, built on Crab, requires a particle conversion efficiency which is an order of magnitude higher than Model 1. Our results are also in agreement with a smooth change in the spectral shape as reported by AMS-02.\\
\indent Comparison with data, for energy above 100 GeV (see solid lines in Fig. \ref{fig:All_LIS10TeVall}), indicates the possibility of an extra source similar to Vela. There is only another known pulsar, B1737-30 (or J1740-3015), with parameters similar to Vela. It is 400 pc far and 20600 years old \citep{Clifton1986,Yuan2010}, but the photon emission of its nebula has not been observed yet. Moreover, statistically we expect just one (or very few) more pulsar like Vela: we can consider a pulsar birth rate of 0.9-2 objects per century \citep{Taani2012,Duncan2008}, a spatial distribution like in \citet{Taani2012} and \citet{Sartore2010}, pulsars with age from 10 to 50 kyrs in a volume of about 1 kpc$^3$ around the Earth.\\
\indent At energies lower than 100 GeV, the tiny discrepancy of Fig. \ref{fig:All_LIS10TeVall} (Model 2) can be covered by mature pulsars ($10^5$ years) because the low energy electrons and positrons released are still diffusing to the Solar System. The main contribution from this sources comes from Monogem and it is a factor 2 higher than the Vela-X one at 40 GeV.\\
\indent Our models predict that a single close PWN, Vela-X, may be responsible for at least half of the electron and positron excess in CRs, thus, a dipole signal in the CR arrival directions could be detected. If no other source is missed, we expect a dipole anisotropy above 100 GeV centered in the direction of Vela. At 200 GeV, the dipole anisotropy for electrons plus positrons from Vela-X is expected to be of the order of $\sim2\%$ \citep{RozzaICRC2015} not yet excluded by the FERMI experiment \citep{Ackermann2010}. Conversely, at lower energy, several sources can contribute to the electron and positron spectra, but the angular distribution of all these sources should be more isotropic. AMS-02 observation could be able to provide additional data to the anisotropy studies.\\
\indent Our models predict a change of the slope at energies below 10 TeV (see Fig. \ref{Fig_PSR}). Future measurements up to this energy range can be used to confirm or reject such a prediction.\\
\indent Our results are in agreement with models describing the origin of the pulsar wind nebula (e.g., the one discussed in \citealt{Blasi2010}), where the minimal electron and positron luminosity is not expected to be lower than a few percent. Thus, in this context, the current results can be used to constrain the fraction of the spin-down luminosity which is transferred to particle acceleration needed to fit the excess spectra observed by AMS-02.\\
\indent Finally, the satisfactory agreement between the models and the data leads to keep into account PWN as source of electrons and positrons. Therefore, a realistic LIS should include this type of electrons and positrons sources.


\begin{thebibliography}{10}
\bibitem{RozzaICRC2015}
D.Rozza, et~al., {Vela-X as main contributor to the electron and positron
  spectra for energy above 100 GeV, to be presented at the 34th ICRC
  conference}.
\end{thebibliography}

\end{document}